\documentclass[12pt,a4paper]{article}

\usepackage{epsfig}
\usepackage{amssymb}

\newcommand{\ncmd}{\newcommand}

\newtheorem{defi}{Definition}
\newtheorem{theo}{Theorem}
\newtheorem{prop}{Proposition}
\newtheorem{lem}{Lemma}
\newtheorem{cor}{Corollary}

\newtheorem{rem}{Remark}

\ncmd{\btheo}{\begin{theo}$\!\!\!$. -- }
\ncmd{\etheo}{\end{theo}}
\ncmd{\bpro}{\begin{prop}$\!\!\!$. -- }
\ncmd{\epro}{\end{prop}}
\ncmd{\preuve}{{\sc Preuve --}\ }
\ncmd{\bdefi}{\begin{defi} $\!\!\!$. -- }
\ncmd{\edefi}{\end{defi}}
\ncmd{\bco}{\begin{cor}$\!\!\!$. -- }
\ncmd{\eco}{\end{cor}}
\ncmd{\ble}{\begin{lem}$\!\!\!$. -- }
\ncmd{\ele}{\end{lem}}
\ncmd{\bno}{\begin{notation}$\!\!\!\!\!$. -- }
\ncmd{\eno}{\end{notation}}
\ncmd{\bre}{\begin{rem}$\!\!\!$. --  \begin{em}}
\ncmd{\ere}{\end{em} \end{rem}}
\ncmd{\beq}{\begin{equation}}
\ncmd{\eeq}{\end{equation}}
\ncmd{\ben}{\begin{enumerate}}
\ncmd{\een}{\end{enumerate}}
\ncmd{\bit}{\begin{itemize}}
\ncmd{\eit}{\end{itemize}}
\ncmd{\refp}[1]{(\ref{#1})}

\ncmd{\Fi}{\mathbb{F}}
\ncmd{\Oc}{\mathbb{O}}
\ncmd{\Ha}{\mathbb{H}}
\ncmd{\R}{\mathbb{R}}
\ncmd{\C}{\mathbb{C}}
\ncmd{\Z}{\mathbb{Z}}
\ncmd{\N}{\mathbb{N}}
\ncmd{\Sph}{\mathbb{S}}
\ncmd{\T}{\mathbb{T}}
\ncmd{\D}{\mathbb{D}}
\ncmd{\Lp}{\mathfrak{p}}
\ncmd{\Lg}{\mathfrak{g}}
\ncmd{\La}{\mathfrak{a}}
\ncmd{\Lk}{\mathfrak{k}}
\ncmd{\Lm}{\mathfrak{m}}
\ncmd{\Lh}{\mathfrak{h}}
\ncmd{\ad}{\mbox{ad}}
\ncmd{\Ad}{\mbox{Ad}}
\ncmd{\ovr}{\overrightarrow}

\ncmd{\di}{\displaystyle}
\ncmd{\bs}{\backslash}
\ncmd{\ov}{\overline}
\ncmd{\no}{\noindent}
\ncmd{\ra}{\rightarrow}
\ncmd{\lra}{\longrightarrow}
\ncmd{\eps}{\epsilon}

\ncmd{\scalar}[2]{\mbox{$\mathcal{h} #1,#2 \mathcal{i}$}}
\ncmd{\ichap}{\^{\i}}

\title{Huygens-Fresnel Principle in Superspace}
\author{H. de A. Gomes\footnote{University of S\~ao Paulo, Institute of Mathematics and Statistics, gomes.ha@gmail.com, hadgomes@ime.usp.br
}}

\begin{document}
\maketitle

\begin{abstract}
 We first roughly present a summary of the optico-mechanical analogy, which has always been so profitable in physics. Then we put forward a geometrodynamical formulation of gravity suitable to our intentions, both formally and conceptually. We present difficulties in some approaches to canonically quantize gravity which can be amended by the idea put forward in this paper, which we introduce in the last section. It consists basically in trying to find an intermediary between the quantization step going from the classical superhamiltonian constraint to the Wheeler-DeWitt equation. This is accomplished by inputting interference beyond the WKB approximation, through a type of Huygens-Fresnel Principle (HFP) in superspace.

\end{abstract}

\section{Introduction}

\subsection{Motivation of the analogy}

In classical particle physics, the Hamilton-Jacobi equation for a fixed energy has an exact counterpart in optics, in the form of Hamilton's partial differential equation for geometrical optics (which represents Huygens' principle in infinitesimal form). As is well known, Huygens' principle, and hence Hamilton's equation, does not account for interference, for that we must go on to HFP or to the full wave equation. The transposition of optical formalism to explain mechanical phenomena is widely known for it's greatest achievement; keeping the same principles of conversion, we get the closest thing there is to a ``derivation'' of the Schroedinger time-independent equation from the D'Alembert wave equation.

Actually, we can extend this parallelism to quantum gravity. The fixed energy Hamilton-Jacobi equation is very similar in form to the superhamiltonian constraint, and so is the time independent Schroedinger equation to the Wheeler-DeWitt (WDW) equation, and the process of going from one to the other follows the same principle of conversion. But what happens is that the WDW equation is unwieldily, and it would be profitable if there was an intermediate approximation in superspace as there is in classical optics.

However, we need not go so far as quantum geometrodynamics to appreciate the usefulness of the HFP. As we will show, by removing a few of the restrictions in the HFP, it can be likened to the construction of the Feynman path integral \cite{Fe}. Of course, the path integral per se does not need the concept of classical trajectory \footnote{In fact one of it's most beautiful features is the way it derives the classical trajectory following very simple first principles, i.e.: the extremalization of the action by coherent summing of phase angles.}, whereas HFP rests on the very thing. In counterpoint, as we will show, in this extended HFP formalism there is no need to presuppose a phase proportional to the action, furthermore we can actually drop the geodesic principle and heuristically proceed in the same manner as Feynman, deriving an approximation to the path integral formalism by taking account not only of the subsystem in question, but of the whole Universe, in step with the whole Machian ideal and machinery of our approach. The added benefit here would be to have an easier to handle measure in superspace and lower order approximation.

\subsection{A Brief History of The Optico-Mechanical Analogy}
Right from the dawn of the physical sciences,
when methods of higher mathematics were still rudimentary, physics gained insight into natural phenomena from the comparison of optical and mechanical dynamics. John Bernoulli, at the XVIIth century,  already treated the motion of a particle in the field of gravity as an optical problem, assuming a refractive index proportional to $\sqrt{E-V}$.

 It is as well a startling fact that Hamilton obtained the basic partial differential equation for geometrical optics (which in fact expresses Huygens' principle in infinitesimal form):
\beq\label{Huygens}{\left(\frac{\partial{\sigma}}{\partial{x}}\right)^2 + \left(\frac{\partial{\sigma}}{\partial{y}}\right)^2 + \left(\frac{\partial{\sigma}}{\partial{z}}\right)^2 = \frac{n^2}{c^2}}\eeq  prior to its application in mechanics (\cite{La}). In (\ref{Huygens}), $\sigma(x,y,z)$ is the time that light requires to travel from a basic surface given by $\sigma=0$ to the given point $(x,y,z)$, and it contains the information that light rays are orthogonal to these surfaces. The analogous to (\ref{Huygens}) in mechanics assumes one of the forms of the Hamilton-Jacobi partial differential equation (for a given energy):
\beq\label{Hamilton}{H\left(q^1,...,q^n,\frac{\partial{S}}{\partial{q^1}},...,\frac{\partial{S}}{\partial{q^n}}\right)=H\left(q^r,\frac{\partial{S}}{\partial{q^r}}\right)=E}\eeq
where H is the Hamiltonian function of the system, $q^i$ are generalized coordinates for the n particles, $E$ the energy, and S is the action, or Hamilton's characteristic function\footnote{Not to be confused with Hamilton's principal function, W:  if we denote the initial points by $\{q_{a}^r\}$ and the final by $\{q_{f}^r\}$ we have that $S(q_{a}^r,q_{f}^r,E)$ while $W(q_{a}^r,q_{f}^r,t)$. While they are easily transformed into one another,and both yield the correct equations of motion from their versions of Hamilton-Jacobi's equation, ours is more appropriate since we want to fix the energy and exclude time.} which obeys \beq\label{momentum}{p_i=\frac{\partial{S}}{\partial{q^i}}}\eeq

If we are dealing with the motion of a single particle in a field of potential $V$, employing regular Euclidean coordinates the energy equation is: \beq{\frac{1}{2m}(p_1^2+p_2^2+p_3^2)+V(x,y,z)=E}\eeq or, in more appropriate form:
\beq\label{energy}{\left(\frac{\partial{S}}{\partial{x}}\right)^2+\left(\frac{\partial{S}}{\partial{y}}\right)^2+\left(\frac{\partial{S}}{\partial{z}}\right)^2=2m(E-V)}\eeq

If, in a conservative system, we start from any given point in the configuration space (which is just $\R^3$ for our single particle) for which we define $S=0$ and assume $\nabla{S}\neq{0}$ \footnote{Actually we may start from any given surface for which $S=0$, but then the ray property would be valid only if the system is conservative and for those paths that start with the same energy and perpendicular to the given surface. All of these restrictions will be removed once we include time in the same footing as the other variables and consider orthogonality in the sense of an intrinsic configuration space geometry.}, we can construct a function S which satisfies (\ref{energy}) and fills a finite portion of Euclidean space with surfaces $S=cte$. We can do that because S is non-singular in a neighborhood around the initial point, and hence inverses of it's regular values form  2-dimensional manifolds.

 Furthermore we get by (\ref{momentum}) that $\nabla{S}=\bf{p}$ , i.e.: the possible mechanical paths are orthogonal to the surfaces $S=cte$\footnote{Since in the Euclidean metric we can identify vectors with their duals, that is $\nabla{S}=dS$} , that is, they behave exactly like light rays of optics with respect to surfaces of equal time, hence equation (\ref{energy}) can be considered as an analogue of Huygens' principle.

In the same manner there is a striking similarity in the way we define both trajectories through minimum principles. Fermat's principle, minimizing the integral:
\beq\label{Fermat}{\int^{\tau_2}_{\tau_1}\frac{ds}{v}=\int^{\tau_2}_{\tau_1}\frac{n}{c}ds}\eeq or, making the substitution $n/c=\sqrt{2m(E-V)}$ we get:
\beq\label{Jacobi}{A=\int^{\tau_2}_{\tau_1}\sqrt{2m(E-V)}ds}\eeq minimizing this integral is nothing but Jacobi's principle, which, fixing the energy (i.e.: for a closed system), is equivalent to the principle of least action. It holds valid for any number of particles.

SO we will drop the restriction on the number of particles, and also truly include time with the other variables, $t\rightarrow~q_{n+1}=\tau$. We arrive naturally at the ``special'' line element: \beq\label{line}{dl=L\left(q_1,...,\tau,\frac{dq_1}{d\tau},...\frac{dq_n}{d\tau}\right)d\tau}\eeq where the lagrangian is supposed to be homogeneous of first order in the velocities. This gives an intrinsic geometry to the configuration space, where dynamical orbits are now geodesic paths. For example, in Newtonian dynamics, the timeless Jacobian line element assumes the form: $dl=\sqrt{(E-V)T}d\tau$   \cite{Ba.1},\cite{La}. It is of course linear and homogeneous in the velocities, since $$T=\frac{1}{2}\sum_{i=1}^nm_i\frac{dx^i}{d\tau}\frac{dx^i}{d\tau}$$ and it can be naturally obtained from (\ref{Jacobi}) by putting $ds=\sqrt{T}d\tau$ since $ds=\sqrt{\sum{dq^idq^i}}$. Equation (\ref{energy}) is the Hamiltonian equation obtained from this Jacobi action for a single particle, but using an intrinsic line element we can generalize the form of (\ref{energy}) for arbitrary coordinates and number of particles.

 If we consider the above mentioned intrinsic line element, we equate the problem of solving the equations of dynamics with that of finding the geodesics in a certain (generally non-riemannian) manifold \cite{La}.  It is here important to notice that this principle is truly timeless, it only yields paths in the configuration space, not the speed with which the system traverses it, that is, it gives importance solely to the relative configuration of the particles, we will comment on this property later.

In this new metric space, the dynamical phenomena corresponds to the propagation of light in an optically homogeneous medium! That is, the mechanical paths are geodesics, the elementary waves in Huygens construction are locally spheres\footnote{ More precisely, it yields geodesic spheres \cite{Bu}, or spheres embedded in the intrinsic geometry of the configuration space}, and their envelopes, the wave surfaces, are parallel surfaces. Hence we find that up to now we have accomplished a complete congruence between optical and mechanical dynamics!

In optics, if we are dealing with a wave of a definite frequency $\nu$ we are allowed to make the substitution $\phi=\nu\sigma$ , where $\phi$ is the phase angle. Then (\ref{Huygens}) becomes:\beq{\left|\nabla\phi\right|^2=\nu^{2}n^2/c^2=1/\lambda^2}\eeq  where $\lambda$ denotes wavelength. Using in (9) a proportional correspondence between action and phase angle to the one we have explored so far, $\phi=S/\hbar$, where $\hbar$ is a constant, we get the deBroglie wavelength \beq{\frac{1}{\lambda^2}=\frac{2m(E-V)}{\hbar^2}=\frac{m^2v^2}{\hbar^2}}\eeq Thus $\lambda=\hbar/mv$.

But of course Huygens' principle is only an approximation of physical optics and wave propagation in general. For instance, up to now we have not assumed any principle of interference between trajectories (which is accomplished for example by the Huygens-Fresnel principle) and this is one of the reasons our mechanical and geometrical optical systems behave almost exactly alike. And of course ultimately we should really consider the full wave equation for the optical field \beq\label{wave}{\nabla^2\Psi-\frac{n^2}{c^2}\frac{\partial^2\Psi}{\partial{t}^2}=0}\eeq If we assume that the optical vibration occurs with a definite frequency $\nu$ we can write $\Psi=e^{2{\pi}i{\nu}t}\psi(x,y,z)$, where $\psi$ is the amplitude function, for which we get the differential equation \beq\label{pre}{\nabla^2\psi+\frac{4\pi^2}{\lambda^2}\psi=0}\eeq If we put in the expression for the deBroglie wavelength in (\ref{pre}) we get Schroedinger's time independent equation\footnote{It is not hard to see that the Hamilton-Jacobi equation can be considered as the first step in the WKB approximation to the Schroedinger equation \cite{Ku}}: \beq\label{sch}{\nabla^2\psi+\frac{8\pi^2m}{h^2}(E-V)\psi=0}\eeq

And of course from here on, with the advent of quantum mechanics and later quantum electrodynamics this dualism coalesced into one, and the analogies were made transparent by  full wave mechanics.

We can also trace out a parallel of this analogy in quantum geometrodynamics (\cite{JW.1}), where the superhamiltonian constraint \beq\label{shc}{G^{ijkl}\frac{{\delta}S}{{\delta}g^{ij}}\frac{{\delta}S}{{\delta}g^{kl}}+R=0}\eeq plays the role of equation (\ref{energy}) and is obtained likewise from a minimum action principle in superspace (or a geodesic principle arising from an appropriate line element linear and homogeneous in the momentum (\cite{JW.2})). We will elucidate the meanings of the terms appearing in this equation in the next section, for now we just want to motivate the connection with the wave-particle analogy.

 We will see that by canonical quantization, (\ref{shc}) gives rise to the Wheeler-DeWitt equation: \beq\label{wdw}{({\hbar}^2G^{ijkl}\frac{{\delta}}{{\delta}g^{ij}}\frac{{\delta}}{{\delta}g^{kl}}+h^{1/2}R)\psi=0}\eeq which plays the part of equation (\ref{sch}), and the analogy stands complete.

However, in optics we may introduce an intermediate step, a combination of Huygens' construction with the principle of interference, called the Huygens-Fresnel Principle, which takes into account some of the vectorial character of light propagation. The aim of this paper is to expound this analogy as a means to dodge some of the problems with the ``full wave equation'' represented by the Wheeler-DeWitt equation.

\section{Geometrodynamics}
\subsection{Canonical Formulation}

We now turn to a different formulation of general relativity, one more apt to be transformed in a dynamical theory in superspace. We follow the route of the ADM approach \cite{ADM} since it is standard, even though we feel that the geometrodynamical flavor of general relativity as a theory of evolving three geometries in superspace is better represented in \cite{Ba.1},\cite{Ba.2},\cite{JW.1},\cite{JW.2}, which we will comment on afterwards.

The ADM approach begins by slicing a given spacetime manifold M into spacelike hypersurfaces and a timelike direction. It means that we are restricting the topology of M to be of the type $\R\otimes{\Sigma}$. Hence we are discarding spacetimes with rotations and closed timelike curves\footnote{Actually this is not such a great restriction, by a theorem of Geroch \cite{Ge}, if M possesses a Cauchy surface, i.e.: if it is globally hyperbolic, it is also stably causal and a global time function can be chosen so that each surface of constant value is a Cauchy surface (see \cite{Wa} for the precise definitions).}, in accordance with the view expressed by \cite{Ba.1}. We will also restrict our attention to the case of compact $\Sigma$, which are technically simpler and conceptually richer. As Wheeler argues \cite{JW.1}, due to quantum fluctuations it would be a difficult business to define an open and asymptotically flat space in the context of quantum geometrodynamics\footnote{We also have to note the distinction between asymptotically flat in the context of four geometries and three geometries. For example, in the Schwarzschild 4-geometry, the mass of the center of attraction is uniquely determined by the rate of approach to flatness, but if we slice spacetime by spacelike three geometries the analogous calculation of the apparent mass depends on the choice of slice \cite{JW.1}}. Furthermore, as a consequence of Thurston's conjecture (now Perelman's Theorem) every simply connected smooth compact three manifold is homeomorphic to $S^3$, and we get sound classificatory theorems\cite{Ch} for our possible $\Sigma$s. For physical reasons we will also consider our manifolds to have no boundary.

Now, each of these constant time hypersurfaces $\Sigma_t$ is equipped with coordinates \{$x^i$\} and a three metric $h_{ij}$ with determinant $h$ and inverse $h^{ij}$. When it is necessary, we'll write the four metric as $g_{\mu\nu}$. But we will write R for the {\it{3-scalar curvature}} and $~^{(4)}R$ for the respective four scalar. We will also use boldface signifying the tensorial quantity usually associated with the symbol in question.

 Our objective is to orthogonally decompose a given vector field in respect to the surface, so that we can cleanly separate time evolution from surface elements. So basically we have to see how the metric is rewritten in the appropriate change of basis. The spatial component really belongs to the surface so there is nothing to do there. However we must decompose the temporal unit vector (which connects points with the same space label $\{x^i\}$) orthogonally to the surface. The tangential projection is called the shift vector, with components $N^i$, and the normal projection is the lapse function $N$. That is $({\bf{v}},t)\rightarrow({\bf{v}}+{\bf{N}}t,Nt)$. Here the boldface ${\bf{N}}$ represents the shift vector. Hence in a general displacement $(dx^i,dt)$ we would get in the new ``orthogonal'' coordinates $(dx^i+N^idt,Ndt)$ ,
and our total displacement would be:\beq\label{ds}{ds^2=-N^2dt^2+h_{ij}(dx^i+N^idt)(dx^j+N^jdt)}\eeq
from the Lorentzian version of Pythagoras theorem.

It is easy to see from our normal decomposition that $\bf{g}=h-n\otimes{n}$ where $\bf{n}$ is a unit normal to the hypersurface.

Since we will treat {\bf{h}} as the true intrinsic metric, we will use it, as opposed to the complete four metric, to lower and raise indices of spatial tensors. From (\ref{ds}) we can see that for an observer with four velocity normal to the hypersurface \beq\label{dtau}d\tau=Ndt\eeq

Now we have to express the Einstein-Hilbert action \beq{S[{\bf{g}}]=\int_M\sqrt{-g}~^{(4)}Rd\bf\nu}\eeq in 3+1 form, where $d\nu$ is the volume element, $^{(4)}R$ is the 4-dimensional scalar curvature, we are ignoring the cosmological constant term and the brackets signify functional dependence. The extrinsic curvature tensor $\bf{K}$ is the symmetric two form ${\bf{g}}(\nabla^T\bf{n},~\cdot~)$ where the T superscript means we are restricting the covariant derivative to be along hypersurface vectors. Since $\bf{n}$ is a unit vector and the covariant derivative commutes with the metric we get ${\bf{g}}(\nabla^T\bf{n},{\bf{n}})=0$, hence this tensor is intrinsic to the surface.

Now, from a rather straightforward but tedious calculation, dropping divergence terms (see appendix E.2 of \cite{Wa}) we get: \beq{\sqrt{-g}~^{(4)}R=\sqrt{h}N(R+K_{ij}K^{ij}-(tr{\bf{K}})^2)}\eeq
And now our action becomes \beq\label{action with momentum}{S[h_{ij},N,N^i]=\int{dt}\int_\Sigma{\sqrt{h}N(R+K_{ij}K^{ij}-(tr{\bf{K}})^2)d\Sigma}}\eeq
And we have a well defined lagrangian, which is just the space integral in the last equation. We can write the rate of change of the metric with respect to the label time as \beq\label{mv}{\frac{1}{2}\partial_t{\bf{h}}=-N{\bf{K}}+\mathfrak{L}_{\bf{N}}\bf{h}}\eeq This is valid in an arbitrary spacetime, but the dynamics is encoded in the second time derivative of the metric, which is to be determined from the variation of the action. And so we find the geometrodynamical momentum by \beq{\frac{\delta{S}}{\delta{h^{ij}}}=\frac{\delta{L}}{\delta{(\partial_th^{ij}})}=\pi^{ij}}\eeq which in terms of the extrinsic curvature becomes $\pi^{ij}=\sqrt{h}(K^{ij}-h^{ij}K)$.

The ADM action is to be varied with respect to the lapse, shift and the spatial metric. We note that the time derivatives of the shift and lapse do not enter into the ADM action, hence they are not dynamical variables, but Lagrange multipliers of constraints, which constrain the way our hypersurfaces are to be laid out in spacetime. Therefore we shall only perform the Legendre dual transformations on the metric velocities and leave the constraints as they are, giving: \beq{S[{h_{ij},\pi^{ij},N^i,}N]=\int{dt}\int(\pi^{ij}\partial_tg_{ij}-N\mbox{H}-N_i\mbox{H}^i)d\Sigma}\eeq
Where H, called the superhamiltonian takes the form: \beq\label{sh}{\mbox{H}=G_{ijkl}\pi^{ij}\pi^{kl}-\sqrt{h}R}\eeq where $G_{ijkl}=\frac{1}{2}{h}^{-1/2}(h_{ik}h_{jl}+h_{il}h_{jk}-\frac{1}{2}h_{ij}h_{kl})$ is the DeWitt supermetric with inverse $G^{ijkl}=\sqrt{h}(h^{ik}h^{jl}-h^{ij}h^{kl})$. And $\mbox{H}^i$, is given by: \beq\label{sm}{\mbox{H}^i=-2\nabla_i\pi^{ij}}\eeq
If we vary the action with respect to the momentum and metric we get the evolution (or Euler-Lagrange) equations for geometrodynamics, and by varying the lapse and shift we get the supermomentum and superhamiltonian constraints $\mbox{H}=\mbox{H}^i=0$, the Hamilton-Jacobi-Einstein equations, which correspond to the Gauss-Codazzi constraint equations for isometric embeddings.

The supermomentum and superhamiltonian constraints are propagated by the evolution equations (this follows from the Bianchi Identity \cite{Wa}), but that is not all, they actually can be seen to play a double role \cite{Ku.2}: they restrict the canonical data on an initial hypersurface, and as dynamical variables, evolve the same canonical data\footnote{This is related to the Interconnection Theorems \cite{Ku}, which use Lorentzian embeddability and the global satisfaction of one  constraint to derive global from local satisfaction of the other}. The Poisson bracket of the data with the supermomentum generates their change by a Lie derivative in the direction along the surface, and the Poisson bracket with the superhamiltonian generates change by a Lie derivative in the direction normal to the surface. That is, on-shell these constraints are the generators of spatial diffeomorphisms (supermomentum constraint) and foliation invariance (superhamiltonian constraint). This is an instance of the problem of time, which mixes constraints with dynamics \cite{Ca}.

  Thanks to the work of  York,  Choquet-Bruhat, and  Lichnerowicz among others, G.R. has been show in this formalism to have a well posed initial value problem \cite{Yo},\cite{Wa}. That is, given the initial constrained canonical data $(\Sigma, {\bf{h},K})$, there exists a unique globally hyperbolic Lorentzian spacetime $(M,\bf{g})$ which satisfies Einstein's equation and induces the three metric {\bf{h}} in a Cauchy surface diffeomorphic to $\Sigma$ with extrinsic curvature {\bf{K}}.

We have so far, been discussing solely source free general relativity. If we do have matter sources we have to modify the supermomentum and superhamiltonian constraints in the following fashion \cite{Wa}: \beq{G_{ijkl}\pi^{ij}\pi^{kl}-\sqrt{h}R=16\pi\rho}\eeq where $\rho$ is the energy density, and \beq{-2\nabla_i\pi^{ij}=8\pi{J}^i}\eeq where $J^i$ is the density of energy flow. For the Einstein-Maxwell and Einstein-Klein-Gordon equations we retain well-posedness of the initial value problem. Furthermore, since for non-derivative couplings the new supermetric is just the direct sum of the DeWitt supermetric with the metric for the coupled fields \cite{Ku}, our results will remain valid for these cases.

\subsection{Timeless Formulation}

Unlike the usual geometrodynamical quantization, a Huygens-Fresnel principle in superspace will not involve any elements off dynamical orbits, hence we can impose constraints that are propagated by the evolution equations that will facilitate our treatment.

In this sense, the treatment that begs to be used is York's ``extrinsic time'' approach \cite{Yo}, which conformally rescales the metric decoupling the constraints (\ref{sh}) and (\ref{sm}) and imposing a time-slicing by constant mean curvature surfaces. This has the advantage of being the only known effective method of solving the initial value problem, but it is also the one that lends itself most easily to a reduced phase space quantization method \cite{Ku}, \cite{Mon}, \cite{Mon.2}, \cite{Ca}. It turns out to be even more interesting since it is possible to derive the same results of this treatment from basic principles, and perhaps get rid of foliation invariance nuisances in quantization, in this we follow the work of Barbour \cite{Ba.1}, \cite{Ba.2}, \cite{RWR}. Hence in our approach we won't have to deal with two of the main disadvantages of this approach, which are the apparent arbitrariness of the gauge-fixing, and the discomfort of discarding off-shell paths\footnote{DeWitt was against relating dynamical paths to the term ``on-shell'', since in this context they had nothing to do with mass, and insisted on calling them dynamical orbits. We will interchange both practices.}.

By a Timeless formulation, we mean one in which each instant is represented by a three dimensional Riemannian geometry, not formerly embedded in a Lorentzian manifold. The space of possible instants is still superspace, but now we don't presuppose any other structure; all the physical information is contained in the intrinsic relative configuration of the instants. We aim to construct a fully Machian (since external reference frames are to be eliminated),  and  Jacobi type formulation. That is, it must yield only paths in configuration space, and not their velocities traversing it, which would have no meaning.
It would have no meaning because an ``instant'' is seen as one configuration of the entire system, and any velocity whatsoever is measured by changes in relative configurations of objects within the system itself. A path in configuration space would contain the motion of all objects in the system, so if we speed up the velocity with which we traverse the path in configuration space, the clocks used to measure speed will be speeded up as much as the motion it measures.

 The aim of this approach is to argue that physics does not need a time dimension, that the entire content of physics can be built up using only relative configurational instants and the intrinsic difference between their concrete contents. How to arrive at the notion of intrinsic difference between configurations will be one of our first concerns. First, we have to set the proper stage.

We haven't, till now, given any attention to the configuration space of geometrodynamics, which is superspace, the space of all Riemannian geometries on the topological manifold $\Sigma$ (which we are considering to be compact without boundary), more precisely $Riem(\Sigma)/Diff(\Sigma)$. That is, each point in superspace is an equivalence class of Riemannian metrics under diffeomorphisms (since they represent the same physical configuration). It has been shown as early as \cite{St} that superspace constitutes a proper manifold, with Hausdorff topology and it is locally homeomorphic to an open set in a Banach space \footnote{Diff acts properly on Riem}. If we took the metric of $\Sigma$ not to be positive definite though, we would not have Hausdorff topology.

The supermomentum constraint ensures this formalism makes sense in superspace, and could be eliminated had we any means to generally work directly with superspace.
We will now quickly develop the BSW formalism, which is more suited to accommodate our principles, and, following the spirit of Wheeler, henceforth shift our attention to the dynamics in superspace per se,  using the DeWitt supermetric to lower and raise indices. For example, we get from (\ref{mv}) and the expression for the metric momentum that \beq\label{mom}{\pi^{ij}=G^{ijkl}K_{kl}=\frac{1}{N}G^{ijkl}(\partial_th_{kl}-\mathfrak{L}_{\bf{N}}h_{kl})=\frac{1}{N}G^{ijkl}\left(\frac{\partial{h_{kl}}}{\partial{t}}-\nabla_kN^l+\nabla_lN^k\right)}\eeq
Now, the transition made by BSW \cite{JW.2} from the ADM hamiltonian (19) is trivial: they first made the replacement $$K_{ij}=-\frac{1}{2N}(\partial_th_{ij}-\mathfrak{L}_{\bf{N}}h_{ij})\rightarrow{k_{ij}}= \partial_th_{ij}-\mathfrak{L}_{\bf{N}}h_{ij}$$ which is the unnormalized normal derivative, to give \beq{S=\int{\sqrt{h}}\left(NR+\frac{1}{4N}(k_{ij}k^{ij})\right)}\eeq remembering that we used the supermetric to raise indices. They then varied the action with respect to the lapse\footnote{Which is a Lagrange multiplier.} and found \beq\label{N}{N=\frac{1}{2}\sqrt{\frac{k^{ij}k_{ij}}{R}}=\frac{1}{2}\sqrt{\frac{T}{R}}}\eeq
For reasons soon to be explained we'll call the quantity $k^{ij}k_{ij}$ the kinetic energy T. If we substitute (\ref{N}) into (\ref{mom}), we get the superhamiltonian constraint, which is here in a sense algebraic instead of variational.

  Substituting the expression for the lapse back in the action we arrive at \beq\label{BSW}{S=\int{dt}\int{d\Sigma}\sqrt{h}\sqrt{R}\sqrt{k^{ij}k_{ij}}=\int{dt}\int{d\Sigma}\sqrt{h}\sqrt{R}\sqrt{T}}\eeq
The Lagrangian density is homogeneous of first order in $k_{ij}$, that is, it is proportional to the square root of a two form on $k_{ij}$, which resembles the Jacobi action in classical mechanics and paves the way for a geodesic principle in superspace, specially if seen on conjunction with the fact that $K_{ij}$ and $h_{ij}$ are enough to determine a dynamical orbit.

Now we will construct general relativity directly from first principles in superspace, and incidentally reach the BSW formalism. We will give meaning to the afore mentioned intrinsic difference between configurations, following \cite{Ba.1}, \cite{RWR} and \cite{Ba.2}.

In superspace, every point has attached to it an entire manifold, and an equivalence class of metrics on top of that. Now generally there is no canonical way to identify two diffeomorphic manifolds, $\Sigma[h_1]$ and $\Sigma[h_2]$, where $h_1$ and $h_2$ represent an entire geometry. In fact there is a high order of infinity of ways to relate said manifolds, no intrinsic sense of identity between arbitrary diffeomorphic manifolds\footnote{Space points have no ``identity'', no intrinsic physical reality attached to them (neither do spacetime points \cite{Ro}). This family of diffeomorphisms is well represented by the concept of active diffeomorphisms, since even if we use tensorial notation, that is, without coordinates, we'll still move the points themselves around going from one manifold to the next.}, which is what best defines them.

However, if we have a well defined distance function between points of Riem, we can establish an equilocality relation used to compare quantities in the two diffeomorphic manifolds by finding the diffeomorphism that induces the minimum distance. Symbolically let us write \beq{d[h_1,h_2]=inf\left\{\int_\Sigma{d\Sigma||h_1-f^*h_2||_{h_1}},f\in{Diff}(\Sigma)\right\}}\eeq Where the subscript$h_1$ is there to remind us that the norm also depends on the metric. We note that in this symbolic form this is not well defined since the volume element also depends on one of the metrics and hence this distance is not symmetric. But it is straightforward to see that this distance function would not depend on the particular representative of the metric: suppose $\tilde{f}$ is the diffeomorphism that actualizes the distance, and $\tilde{h}_1\in[h_1],\tilde{h}_2\in[h_2]$ two different elements of the group orbit, we have that $\tilde{h}_1=\sigma_1^*h_1$ for some $\sigma_1\in{Diff(\Sigma)}$, and the same for $h_2$ hence, using the $\sigma$ induced isometry of the norm $||\sigma_1^*(~\cdot~)||_{\tilde{h}_1}=||~\cdot~||_{h_1}$,  we get that $$d[\tilde{h}_1,\tilde{h}_2]=inf\left\{\int_\Sigma{d\Sigma||\sigma_1^*h_1-f^*\sigma_2^*h_2||_{\tilde{h}_1}},f\in{Diff}(\Sigma)\right\}$${$$=inf\left\{\int_\Sigma{d\Sigma||\sigma_1^*\left(h_1-(\sigma_1^{-1})^*f^*\sigma_2^*h_2\right)||_{\tilde{h}_1}},f\in{Diff}(\Sigma)\right\}=d[h_1,h_2]$$ since the compositions and inverses of diffeomorphisms are diffeomorphisms and the pullback by a diffeomorphism doesn't affect the value of the integral. Furthermore, the difference in the volume element is immaterial in the limit of close-by metrics.

Now, following a path in configuration space will generate a one parameter curve of diffeomorphisms $f_t$ in the above fashion. We assume $f_t$ is differentiable in $t$ and call its generator $\xi(t)$, which can be identified as the vector field that generates the infinitesimal change in location between two nearby metric manifolds. Hence, to see how a quantity evolves along this curve in configuration space we have to take into account the change in location, and in the limit, the time derivative of a function has to be corrected by $\xi$, which is accomplished by the Lie derivative of said quantity in the direction of $\xi$. The process of extremalizing with respect to $\xi$ is called best-matching, or free end-point variation (which it actually is \cite{Ba.2}, \cite{RWR}). So, for example \beq\label{1234567890}{\frac{d{\bf{h}}}{dt}=\frac{\partial{{\bf{h}}}}{\partial{t}}-\mathfrak{L}_\xi{\bf{h}}={\bf{k}}}\eeq
We call attention to the remarkable similarity in concept and form between this best-matching and the time derivative in Euler's formalism of fluid dynamics \cite{Fa}.

We have now seen why we called $k_{ij}k^{ij}$ the kinetic term. It will turn out to be useful later to notice as well that since (using the supermetric to lower and raise indices):
$$k^{ij}=\frac{dg^{ij}}{dt}=\frac{dg^{ij}}{d\tau}\frac{d\tau}{dt}=\frac{dg^{ij}}{d\tau}N$$ and
$$\pi^{ij}=\frac{1}{N}k^{ij}$$ we get that
\beq\label{proper}{\pi^{ij}=\frac{dg^{ij}}{d\tau}}\eeq
This is a very meaningful, completely local equation which we will explore later.

Using the above mentioned kinetic term we see as well that (\ref{BSW}), the BSW action, has the form of such an infinitesimal distance function (modulated by a conformal factor that detrivializes the theory, given by the scalar curvature).

The position of the square-root in (\ref{BSW}) is also of crucial importance, giving local reparametrization invariance of the action with respect to the label time $t$. That means N effectively measures a local rate of change of the metric, homogeneous in $(dt)^{-1}$, hence proper time, given by the relation $d\tau=Ndt$ and equation (\ref{mom}) is reparametrization invariant, i.e.: does not depend on  reparametrizations of the unphysical label time. This is perceived in \cite{Ba.2} as translating a ``Time is derived from change'' principle, or a timeless dynamic in the sense alluded to in the beginning of the section. We also note that, from (\ref{dtau}) and (\ref{N}), regarding $d\lambda$ as the infinitesimal action, or infinitesimal distance between metrics in superspace, we can write the following relation (in B.M. `` coordinates''):
\beq\label{propertime}d\tau=\frac{1}{\sqrt{R}}\sqrt{G^{ijkl}dg_{ij}dg_{kl}}=\frac{||dg||}{\sqrt{R}}
\eeq
So we may regard the passage of proper time (in best-matching coordinates) as a local analogue of the metric change, weighted by the scalar curvature \footnote{If we compare $||dg||$ with momentum attributed to the metric, then $d\tau$ is the speed of the metric change, while $R$ has a role similar to that of a varying mass of the `` particle'' along the trajectory.}.

  Now, $\xi$ can be seen as a gauge auxiliary that renders the action invariant under 3-diffeomorphisms, that is, it only gives sense to motion as an intrinsic change of configuration, it is interpreted as a ``motion is relative'' principle, implementing fully Machian dynamics.

 Generalizations of actions obeying these two principles in superspace were studied in \cite{RWR}, and the requirement of propagation of constraints led almost uniquely back to the BSW action. We note here that there are two extremalizations taking place in this geodesic principle in superspace, the best-matching, that tells us how to compare relative configurations, how to measure their intrinsic distance, and the Jacobi geodesic principle, that chooses the path that minimizes this distance.

 The interesting thing in this approach, is that Barbour et al did not use equation (\ref{ds}) to derive their results, i.e.: they did not use in their construction of general relativity Wheeler's idea \cite{JW.1} that three geometries always evolve in the exact manner necessary to keep embeddability in a four dimensional Lorentzian spacetime \footnote{Wheeler's idea is taken to it's ultimate consequences in \cite{hkt}, where they give a new derivation of GR as a constrained Hamiltonian system}. Even though both can be considered as new derivations of GR, the one given here is more dynamical, without presupposing spacetime at all.

We could extend this line of reasoning, arguing that there  is no sense in attaching meaning to sizes at different points in $\Sigma$. Again, if we change scales of everything going from one point to the next how can we locally perceive it?

Weyl was the first to try to implement some sort of relativity of size principle \cite{We}. In an attempt to unify electricity and gravity he regarded non-metric compatible connections, introducing the electromagnetic vector potential as the source of incompatibility. Unfortunately in this case size turns out to be path dependent, and, as Einstein pointed out, the spectral lines of atomic elements are independent of their past histories. Others tried to pick up where Weyl left off, including Dirac \cite{Di}, and this approach has reached our days in the form of dilaton fields.

 Barbour considered another type of relativity of scale, one that retains metric compatibility of the connection, that is, conformal transformations of the metric. Implementing this principle would mean our theory had to be invariant with respect to conformal transformations.

However, there is one more element we must consider: general conformal transformations alter the spatial volume of the given 3-manifold, but if it is possible to locally, physically detect volume we certainly cannot make the theory invariant with respect to general conformal transformations. Indeed this is the case, at least in compact Riemannian manifolds. There is a theorem from the study of Laplacians on compact Riemannian manifolds \cite{Ri} that states that we can locally detect the volume of the whole manifold, for example from the local solution of the heat distribution (after an infinite amount of time has elapsed).

So Barbour et al \cite{Ba.2}, implementing this further principle of volume-preserving-conformal-transformation (VPCT) invariance, arrived not only at canonical geometrodynamics (with a configuration space consisting of quotiening superspace with respect to VPCTs), but also, if seen extrinsically, at a condition equivalent to a preferred foliation: the constant mean curvature (CMC) foliation utilized by York \cite{Yo} to solve G.R.'s initial value problem! In this approach however, this foliation is not considered as a gauge-fixing condition, as it is by York, but as emerging from first principles of symmetry, and hence it introduces a notion of global simultaneity. In fact, this method does not derive the superhamiltonian constraint explicitly, but an analogous equation that ensures propagation of the CMC condition. That is, we trade the relativity of simultaneity principle (represented by (\ref{shc})) for the relativity of local size principle, we cannot have both at once.

Taking the view from full four dimensional general relativity, such a CMC time slicing of spacetime is conjectured always to exist for spatially compact spacetimes \cite{Ma}, and it amounts physically to using the rate of expansion of the Universe as time. It also must be taken into consideration that this construction gives automatically the right number of degrees of freedom of the gravitational field per space point, 2, represented by the conformal geometry, the shape of space.

 We have gone through the trouble of expounding Barbours's theory not only for the sake of conceptual beauty and simpleness, but also because it is fundamentally connected to one of the major problems in canonically quantizing gravity: the absence of a  phase space for G.R. with only ``true dynamical degrees of freedom'', that is, a configuration space where we need not impose any constraints \cite{Wa}.

\section{Quantum Geometrodynamics}

\subsection{Some Problems in Canonical Quantization}
For a thorough review on the advances, difficulties and extensive bibliography on canonical Q.G. see \cite{Ca}, which we largely follow.

The canonical procedure for quantizing a system was developed by Dirac \cite{Di.2}:
 \begin{enumerate}
\item Take the states of the system to be described by wave functions $\Psi[h_{ij}]$ defined in an auxiliary Hilbert space  $H^{(aux)}$.
\item The Poisson brackets turn into commutators, $$\{h_{ij}(x),\pi^{lk}(x')\}\rightarrow\frac{1}{i\hbar}[h_{ij}(x),\pi^{lk}]=\frac{1}{i\hbar}(\delta_i^k\delta_j^l+\delta_i^l\delta_j^k)\delta(x-x')$$ thus replace each configuration variable by differentiation with respect to the conjugate configuration variable, e.g.: $\pi_{ij}\rightarrow{i}\hbar\frac{\delta}{\delta{h_{ij}}}$

\item Write the constraints as acting on $H^{(aux)}$ and demand that physical states be annihilated by these operators.
\item We form a new Hilbert space,  $H^{(phys)}$ by finding a new inner product on the space of physical states, i.e.: those that are annihilated by the constraints.

\end{enumerate}

Of course the supermomentum and superhamiltonian constraints could not have been supposed to be operator identities since then all commutators with them would have to be zero, and this does not hold true for the classical Poisson bracket analogue. They are first class constraints \cite{Di.2}, and so we impose them on the wave function: \beq\label{wdw}{({\hbar}^2G^{ijkl}\frac{{\delta}}{{\delta}h^{ij}}\frac{{\delta}}{{\delta}h^{kl}}+h^{1/2}R)\Psi=0}\eeq and \beq\label{smc}{2i\nabla_j\left(\frac{\delta\Psi}{\delta{h_{ij}}}\right)=0}\eeq

Every step has it's own difficulties \cite{Ca}, but we will only mention those of the constraint equations (\ref{smc}) and (\ref{wdw}), and of the posterior construction of a Hilbert space.

Equation (\ref{smc}) is easily interpreted, it simply means that the value of the wave function does not change if we take a different representative of the metric, that is, it is diffeomorphism invariant and hence we can take it to be a function on superspace. To sketch a proof of this fact, we make an infinitesimal coordinate transformation, $x^i\rightarrow{x^i+k^i}$ which changes the metric by a Lie derivative $h_{ij}\rightarrow{h}_{ij}+2\nabla_{(i}k_{j)}$. The new wave function may be expanded yielding:

\beq{\Psi[h_{ij}+2\nabla_{(i}k_{j)}]=\Psi[h_{ij}]+\int{2\nabla_{(i}k_{j)}\frac{\delta\Psi}{\delta{h_{ij}}}d\nu}}\eeq

Using integration by parts, canceling boundary terms we arrive at the change in the state function by a coordinate transformation:

\beq{\delta\Psi=-\int{k}_j\nabla_i(\frac{\delta\Psi}{\delta{h_{ij}}})d\nu=0}\eeq

 Equation (\ref{wdw}) is not so tame. Besides operator ordering ambiguities and the lack of a well defined  meaning to the product of two functional derivatives, there are further problems we must confront to make use of (\ref{wdw}).

The DeWitt supermetric is a $6X6$ matrix per space point, it is hyperbolic, with signature $(-,+,+,+,+,+)$ \cite{Ku}, and so it formally resembles a Klein-Gordon equation with variable mass term in superspace. However, attempts to construct a Hilbert space out of the (heuristic) solutions of (\ref{wdw}) following a Klein-Gordon-like construction of the inner product have so far been foiled (\cite{Ku}), and modifications thereof have not fared much better (\cite{Ca}).

The usual path in Klein-Gordon-like theories, is to construct an analogue of the current vector of a relativistic particle in curved space-time, which satisfies a continuity equation. The natural analogue here is: \beq\label{current}{S^{(12)}_{ij}(x)[\Psi_1({\bf{h}}),\Psi_2({\bf{h}})]=\frac{1}{2}G_{ijkl}(x)\left(\Psi_1({\bf{h}})\frac{\delta\Psi_2({\bf{h}})}{\delta{h}_{kl}}-\frac{\delta\Psi_1({\bf{h}})}{\delta{h}_{kl}}\Psi_2({\bf{h}})\right)}\eeq

  where the $\Psi$ are solutions of the WDW equation. If we consider the supermetric compatible connection in superspace $\nabla$, we have that (\ref{current}) obeys the continuity equation:
\beq{\frac{\nabla}{\delta{h_{ij}}}S^{(12)}_{ij}(x)[\Psi_1({\bf{h}}),\Psi_2({\bf{h}})]=0}\eeq
Now, by choosing a ``spacelike'' hypersurface\footnote{Spacelike with respect to the supermetric.} with directed surface element $d\Omega^{ij}$ we can form the Klein-Gordon-like inner product defined by the functional integral:
\beq\label{innerKG}{\scalar{\Psi_1}{\Psi_2}=\prod_{x\in\Sigma}\int_\Omega{d}\Omega^{ij}(x)S^{(12)}_{ij}(x)[\Psi_1({\bf{h}}),\Psi_2({\bf{h}})]}\eeq It can be shown that this does not depend on the choice of the hypersurface \cite{Ku}. As in ordinary Klein-Gordon theory (\ref{innerKG}) is not positive definite, and hence involves unphysical negative probabilities. This is ordinarily solved by restricting to positive energy solutions, which must involve a hypersurface orthogonal conformal ``time-like'' Killing supervector to be defined by. However, the method fails in finding a complex structure $J$, compatible with this inner product, that turns the solution space into a complex vector space, and through which we can find a positive definite inner product \cite{Ku}. The WDW equation is real and hence does not couple real and imaginary parts of a solution in any natural way.

The problem is that here, unlike in the relativistic particle theory, we cannot put the blame of this failure on particle creation, since the WDW equation already is second quantized. In \cite{Ku}, the ``failure in constructing a Hilbert space out of the solutions of the WDW equation'' is seen as a ``failure of equation (\ref{wdw}) itself''.

We note in passing that the Wheeler-DeWitt equation (\ref{wdw}) is quadratic in all momenta, unlike Schroedinger's equation, and this makes it impossible to single out a time variable. In fact, if we make the analogy with the Schroedinger-like functional equation: \beq\label{ssch}{i\hbar\frac{\partial\Psi}{\partial{t}}=\hat{H}\Psi}\eeq
where $\hat{H}$ is the operator correspondent to the classical hamiltonian, imposing (\ref{wdw}) we get that a better analogy  would be made with the Time independent Schroedinger equation. This sits well with the analogy between (\ref{shc}) and (\ref{energy}). There, we used Jacobi's principle, fixing the energy in (\ref{Hamilton}) and (\ref{energy}). As we mentioned before, it is the proper action to be considered for a closed conservative system, and upon quantization it yields the time independent Schroedinger equation. So we have every reason to interpret timelessness of (\ref{wdw}) as signaling a closed system, which is correct, since it is a function of the whole Universe.

Another line of approach is to quantize an already fully constrained system, that is, constrain and then quantize, and so this does not make immediate use of the Wheeler-DeWitt equation. One of the main difficulties here is to solve the classical constraints, i.e.: to find the ``true degrees of freedom''.

One of the most promising approaches is that of Fischer and Moncrief (\cite{Mon}, \cite{Mon.2}), which makes use of York's afore mentioned extrinsic time gauge. As Carlipp objects in his review \cite{Ca} though, in passing to the reduced phase space we freeze many degrees of freedom classically, treating only a subset of physical fields quantum mechanically and maybe excluding important quantum effects. Another objection is that in the Fischer-Moncrief approach we fix time to CMC slices, and so define quantum states only on those slices. But it is not agreed that different time slices will yield the same transition amplitudes between fixed initial and final CMC surfaces. In our approach this last objection is somewhat circumvented, because the time slices were not gauge-fixed.

\subsection{The Huygens-Fresnel Principle}

In Huygens'construction, every point of a wave-front emits secondary spherical wavelets, as if a new disturbance had occurred. This principle, as mentioned in the introduction, is mathematically equivalent to equation (\ref{Huygens}) and so accounts for the characteristics mentioned therein, but as it stands in optics, it is just a mathematical device, having no inherent physical reality. On the other hand, a heuristic analogy between Feynman's path integral and Huygens' Principle in configuration space arises quite naturally: suppose a source is emitting waves with equal amplitudes at all velocities, after a short time interval these wavefronts will be considered as the sources of secondary waves, and so on. If, in the phase, we replace the time by the action, taking infinitesimal time intervals between re-emissions, the result will be that waves can travel from the initial point (at initial time) to a given final point at final time along any path, and the contribution of a given path, the phase,  will be given by the action integral of that path.

It seems worthwhile to mention here that in the HFP in superspace there is no sense in the path velocity, the line element is a super-lagrangian\footnote{Not the super-lagrangian density!} , hence  we  have the simplifications of not having to deal with various velocities and of making the wave number a constant! We must issue a warning: we are not trying to exactly transplant the HFP to superspace, we are using it more as a guideline, as an inspiration of a procedure which might approximate wave behavior (WDW equation) in superspace, since the principle is responsible for approximating it in Euclidean space. In Euclidean space the procedure is already not exact, hence sometimes, in order to better implement characteristics we want in a quantum geometrodynamical theory, we may diverge a bit from the exact translation of classical HFP formalism.

The extension made by Fresnel on Huygens'construction consists of postulating that the secondary wavelets mutually interfere, and it accounts for diffraction and also correctly describes the propagation of light in free space.
We will first follow the usual approach to the HFP, given in \cite{Max}, which we recommend for a complete account since we will only brush over it's main features.

\begin{figure}
\begin{center}
\includegraphics[width=0.4\textwidth,angle=90]{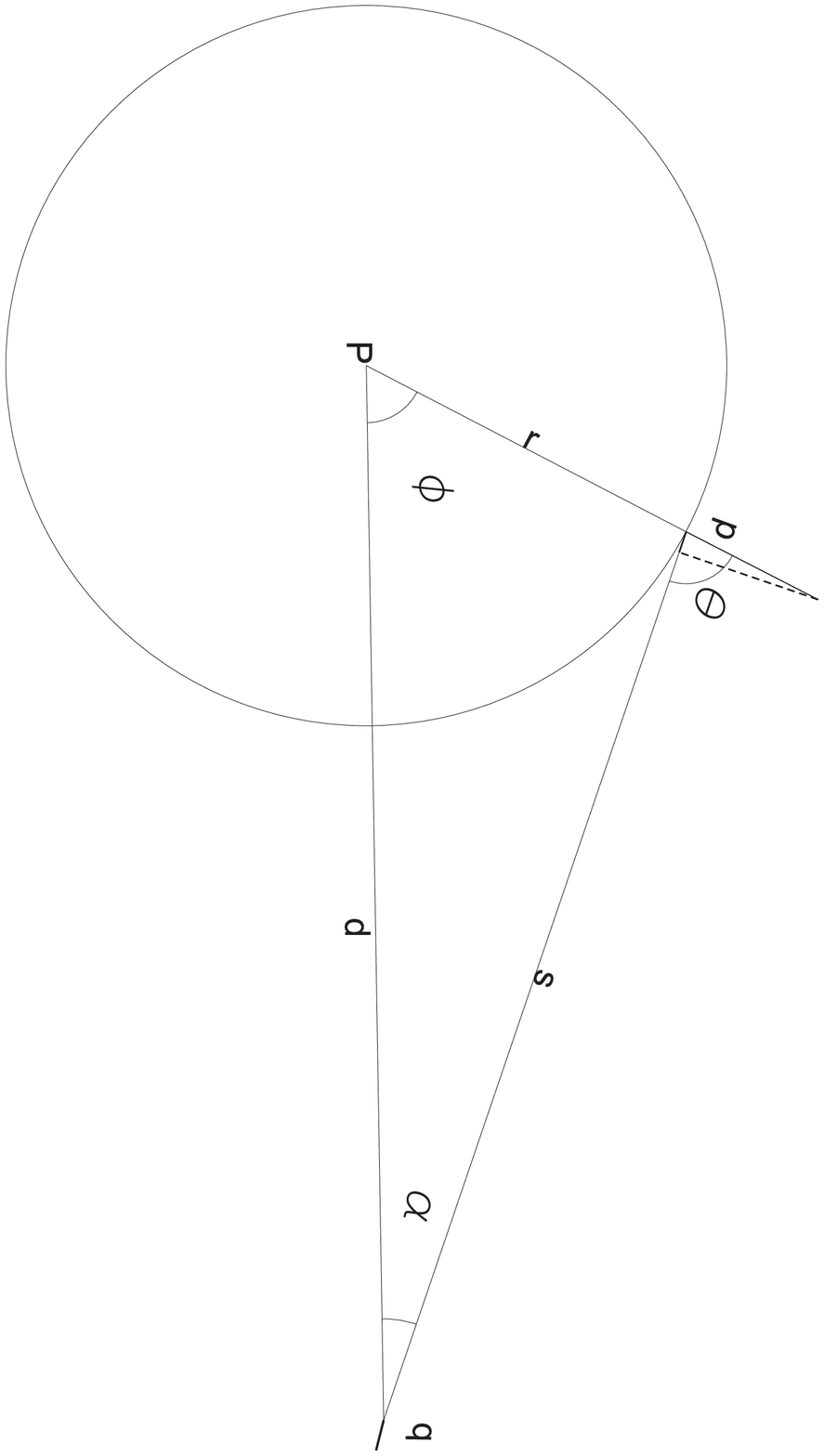}
\caption{Huygens-Fresnel construction}
\label{classical HFP}
\end{center}
\end{figure}

Let $P$ be a point source of a monochromatic wave (see Figure\ref{classical HFP}), if we omit the time periodic factor $e^{i\omega{t}}$, the disturbance at a point $q$, is given by $\frac{A}{r_0}e^{ikr_0}$ , where A is the amplitude a unit distance from the source, $r_0$ is the distance from $P$ to $q$ and $k$ is the wave number. According to the HFP, if we want to determine the light disturbance at a point $p$ further than $r_0$ from $P$, we regard every point of a sphere, labeled $S_1$, centered at $P$ and with radius $r_0$, as the source of a secondary wavelet. The contribution $dU(p)$, due to the infinitesimal area element $dS_1$ situated at $q$, is given by \beq\label{HFP1}{dU(p)=K(\theta)\frac{Ae^{ikr_0}}{r_0}\frac{e^{iks}}{s}dS}\eeq  where $s$ is the distance between $p$ and $q$ and $K(\theta)$ is an inclination factor which describes the variation of the amplitude in terms of the angle $\theta$ between the direction of reemission and the radial vector (i.e.: the original direction of propagation). So the total disturbance at $p$ is given by \beq\label{HFP2}{U(p)=\frac{Ae^{ikr_0}}{r_0}\int_{S_1}\frac{e^{iks}}{s}K(\theta)dS_1}\eeq

From the form of (\ref{HFP2}), if we replace $s$ by the action \footnote{Or, in superspace keep it as distance.} and forego of $K(\theta)$, it is apparent that there {\it{should}} be an analogy between the path integral and Huygens'principle.

The construction goes on to find out the approximate form of $K(\theta)$ and to show how the HFP construction is just really an approximation of the Helmholtz-Kirchoff integral theorem, which expresses the solution of the homogeneous wave equation in an arbitrary point in terms of the value of the solution and it's first derivatives on an arbitrary surface surrounding said point. By later comparing both formalisms, approximating quite a bit, they find \beq\label{K}{K(\theta)=-\frac{i}{2\lambda}\left(1+cos(\theta)\right)}\eeq   which correctly describes the propagation of light in free space.

First of all, would we still get the right solution for $q$ if we, using the same rule, put in another spherical shell, $S_2$ ,between $q$ and $S_1$?  We would get the right solution for $S_1$, since the field at each point in $S_1$, as the sum of the second wavelets emitted by $S_2$, would have the correct value. When $S_1$ ``reemits'' towards $q$ however, we have to redefine $\theta$ along $S_1$ as the angle of emission relative to the normal of $S_1$. The angle cannot be defined relative to a incoming ray,  since now we have an infinity of incoming directions at each point of $S_1$.  Here we have an explicit analogy, let $\Psi(p)$ be the field at the initial point $p$, then to find out the field at the final point $\Psi(q)$ made by $\Psi(p)$ (see Figure\ref{Feynman}): \beq\label{comm}{\langle{\Psi(q)}|{\Psi_1(p)}\rangle=\int_{S_n}...\int_{S_1}\langle{\Psi(q)}|{\Psi(x_n)}\rangle\langle{\Psi(x_n)}|\Psi(x_{n-1})\rangle...\langle{\Psi(x_1)}|{\Psi(p)}\rangle{d}x^i..dx^n}\eeq
So the answer is obviously yes, we would get the correct field for $q$.

Suppose now we would like to transplant this formalism to a WKB approximation of the wave function in superspace. We have to transform $\cos \theta$ into an inner product in superspace, between the tangent vector to the `` geodesic'' connecting an initial point $[g]$ to a point $[h]$ and the tangent vector of the geodesic connecting $[h]$ and the second  metric $[f]$. To be more precise, we define
$$
\begin{array}{lll}
E:TM &\ra& M\times M\\
~v&\mapsto&(\pi(p), \exp(v))
\end{array}$$
where $M$ is any riemannian manifold. It is easy to demonstrate that this map is an immersion whenever $\exp$ is nonsingular. And so defining, for $p,q\in U\subset M$ sufficiently close, $\ovr{pq}:=E^{-1}(p,q)$, we have the substitution $$\cos \theta\ra -\langle \frac{\ovr{[g][h]}}{||\ovr{[g][h]}||},\frac{\ovr{[f][h]}}{||\ovr{[f][h]}||}\rangle$$
With due care, the actual transposition of this procedure should yield a better approximation than WKB.

We would like to mention that this procedure arouses another kind of approximation that might also be of some value, since it has a simpler prescription.
In the usual HFP, we have used vectorial character for incoming and outgoing rays, so we might stick with this notion a little and consider the amplitude field as a vector instead of a scalar field. So let us assume that we define the field at the point $\ovr{x}$ with no intervening spheres as $$\ovr{U}(x)=\frac{A}{r_o}e^{ikr_0}\frac{\ovr{x}}{||\ovr{x}||}$$ where obviously $||\ovr{x}||=r_0$ . Its value at $p$ will be similarly defined by
 \begin{eqnarray*}
 \overrightarrow{U}(p) &=& \int_{S_1}\langle\overrightarrow{U}(x),
 \frac{\ovr{xp}}{||\ovr{xp}||}\rangle \frac{e^{ik||\ovr{xp}||}}{||\ovr{xp}||}\frac{\ovr{xp}}{||\ovr{xp}||} dS_1\\
 ~&=&\frac{Ae^{ikr_0}}{r_0}\int_{S_1}\frac{e^{iks}}{s}(-\sin{\alpha})\cos{\theta}\frac{\ovr{p}}{||\ovr{p}||}dS_1\\
 ~&=& \frac{Ae^{ikr_0}}{r_0}\int_{S_1}\frac{e^{iks}}{s}(s+r\cos{\theta})\frac{\cos{\theta}}{d}\frac{\ovr{p}}{||\ovr{p}||}dS_1
 \end{eqnarray*}
That is, we are defining a mechanism by which the amplitude field is constructed: an amplitude field at point $p$, that is $\Psi(p)$, contributes to the amplitude field at point $q$ with a vector parallel to $\ovr{pq}$ with  magnitude of a point source of amplitude \beq\label{1234567a}\langle\Psi(q),\frac{\ovr{qp}}{||\ovr{qp}||}\rangle\eeq
Of course, in electrodynamics this is  physically absurd, since there is no field in the direction of propagation, i.e.; in the direction of the Poynting vector.

Furthermore we note that such field may be only piecewise smooth, that is, if we take a finite number of spheres between $P$ and $q$ the resulting paths would have a finite number of discontinuities in their derivatives. The amplitude field propagated along a path that is not a broken geodesic is almost always null, since it is the limit of an infinitely broken geodesic, and for each break the probability is  diminished.

\section{Conclusion}
We have constructed an analogy for the Wheeler DeWitt equation of the further approximation that the HFP represents to the usual wave equation. In so doing we have noticed that another perhaps useful calculational tool can be made by a  vectorial HFP.

Now we must verify that an oscillation of the wave function in superspace that is proportional to the superdistance reduces appropriately to give wave numbers in ultralocal space. From equation (\ref{proper}) we get that, if $k$ denotes the ultralocal wave-number, or the phase change per unit ultralocal change of metric\footnote{Here we suppose that $S\left({\bf{g}}(t_0),{\bf{g}}(t_f)\right)$ and $g^{ij}(x)=g^{ij}(x,t_f)$, but we will continue to keep the dependence on the parametrization variable $t$ implicit.} :

\beq\label{wavelength}{k\propto\frac{\delta{S}}{\delta{g}^{ij}(x)}=\pi_{ij}(x)=\frac{dg_{ij}(x)}{d\tau}}\eeq

Hence, since we are assuming that the path's phase (in superspace) is proportional to the action, we get the correct approximate value of the wavelength if along the path's worldline $dg_{ii}\approx{mdx}$, where $dx$ is the ``best-matched''or corrected displacement. This is a pretty reasonable demand, specially if we remember that in the corrected coordinates, dx is a displacement on terms of equilocal points, which are defined by a minimization of global metric variation. So $dg$ along the path of the particle should be proportional to a weighed displacement in equilocal points. It is remarkable that we get a reasonable, completely local phase factor from the global phase.

In fact, this idea of transposing to superspace for better approximations can be extended to other mechanisms for constructing wave character. We give one example that we will expand on  further work:

 Under mild conditions, Riem$(M)$ can be given a natural principal bundle structure, with base space diffeomorphic to superspace $\mathcal{S}$ and with structural group Diff$(M)$. Each $p\in\pi^{-1}([g])$ can be seen as an identification of points over $M$ (NOT coordinates), so there is no canonical identity element of the action of Diff$(M)$. We could take the associated bundle to have typical fiber given by $F=\Gamma(E)$, where $E=\{(x,v), x\in M, v\in \C\}$. A standard representation of Diff$(M)$ on the fiber could be given by
$\rho:G\ra \mbox{Aut}(\Gamma(E))$ such that $(\rho(g)\cdot s)(x)=s(g(x))$. Barbour's best-matching is equivalent to a connection form $\omega$ on $P$. It is strictly analogous to a connection form over a bundle of bases : $\omega$ allows us to determine how much of a given infinitesimal change of bases (an infinitesimal dislocation in $P$) was due to a change of basis. Changing the word `` basis'' to ``identification of points over $M$'' accomplishes the analogy. 

For example, since $\Gamma(TM)$ is the infinitesimal generator of Diff$(M)$, we have that $\omega(v)=\xi\in\Gamma(TM)$. The horizontal projection is naturally given by $v\mapsto v-\omega(v)$, so, since the natural action of $\Gamma(TM)$ on the space of metrics is given by $\xi\mapsto L_\xi$, we recover the best-matching procedure.  There is in fact much work to be done here, and we will present the results soon.

According to our view, in superspace there is no time, instants are represented by metric configurations, and the passage of time is a change in the metric configuration. By (\ref{propertime}) proper time in best matching coordinates can be seen as a weighted local metric change.
  We adopt an Everett interpretation of this formalism, two histories cannot be said to superpose {\it{unless they cross at the same point}}.
That is, paths do not interfere with each other unless they reach the {\it{exact same 3-geometry}}. That is why, for example,  the two-slit experiment, performed with a polarizer in one of the slits, ceases to show interference; the different paths are not reaching the same configuration. According to this view, that is the reason we can never measure anything being simultaneously \footnote{With respect to our CMC foliation, at least! But since this formalism could have been mounted with any consistent spacelike foliation (see Section 2.1), it should be valid regardless thereof.} in two different positions.

Hence when we make a measurement
the effect is not that all these other alternatives (or metrics, or points in superspace) instantaneously cease to exist, but that we weren't on them to begin with. For example, Wigner's friend is {\it{really}} on only one of two different paths\footnote{Actually, two is an idealized quotient of alternatives, as we discussed.} in superspace, but if these paths should ever cross their amplitude fields will interfere. Now it is very important that we remember that for these paths to cross we must have the same identical configurations, same memories, same final elapsed times in all watches, same everything, so no alternative friend will have a different recollection of the path he went through. In these terms, such paradoxes as EPR's cease to exist: We are always on a unique path {\it{through superspace}}. This entails a non-locality of sorts\footnote{In knowledge, not in actual physical variables.}, since if you ultralocally determine which path in superspace you are in, you in principle determine implicitly the configuration at all other points in $\Sigma$.

In superspace, since we are taking into account every spatial object in the universe, and the other degrees of freedom accountable to the geometric field itself, two given classical paths leaving the same initial point diverge very rapidly. By interaction, any minor difference in trajectory is enormously amplified in a short ``time''. That is why it is so difficult to have large scale interference: in superspace, if ``geodesic'' paths  wander off a great distance, the change of direction made so that they can later coincide again will probably diminish the probability, via (\ref{1234567a}}),  by too great an amount.

\end{document}